# BENYO-S2ST-Corpus-1: A Bilingual English-to-Yorùbá Direct Speech-to-Speech Translation Corpus

Emmanuel Adetiba[1,2], Abdultaofeek Abayomi[3], Raymond J. Kala[4],
Ayodele H. Ifijeh[1], Oluwatobi E. Dare[1], Olabode Idowu-Bismark[1], Gabriel O. Sobola[1], Joy N. Adetiba[5], Monsurat Adepeju Lateef[6]

[1]Covenant Applied Informatics and Communications Africa Center of Excellence (CApIC-ACE), Advanced Signal Processing and Machine Learning Research (ASPMIR), & Department of Electrical and Information Engineering, Covenant University, Ota, Nigeria.

[2]HRA, Institute for Systems Science, Durban University of Technology, Durban, South Africa.

[3]Faculty of Engineering, Built Environment and Information Technology, Walter Sisulu University, East London 5200, South Africa & Department of Information Technology, Summit University, Offa 250101, Kwara State, Nigeria.

[4]International University of Grand-Bassam, Grand-Bassam, Côte d'Ivoire.

[5]Department of Nursing, Durban University of Technology, Durban, South Africa.

[6]Faculty of Health Sciences, Durban University of Technology, Durban, South Africa.

## Abstract

There is a major shortage of Speech-to-Speech Translation (S2ST) datasets for high resource-to-low resource language pairs such as English-to-Yorùbá. Thus, in this study, we curated the Bilingual English-to-Yorùbá Speech-to-Speech Translation Corpus Version 1 (BENYO-S2ST-Corpus-1). The corpus is based on a hybrid architecture we developed for large scale direct S2ST corpus creation at reduced cost. To achieve this, we leveraged non speech-to-speech Standard Yorùbá (SY) real-time audios and transcripts in the YORULECT Corpus as well as the corresponding Standard English (SE) transcripts. Notably, the YORULECT Corpus is small scale(1,504) samples and it does not have paired English audios. Therefore, we generated the SE audios using pre-trained AI models (i.e. Facebook MMS). We also developed an audio augmentation algorithm named AcoustAug based on three latent acoustic features (i.e. pitch, volume and speed) to generate augmented audios from the raw audios of the two languages. Based on the augmentation, the BENYO-S2ST-Corpus-1 has 12,032 audio samples per language, which gives a total of 24,064 sample size. The total duration for English audios is 17.81hours whereas for Yorùbá audios, the duration is 23.39hours. Thus, the total audio duration for the two languages is 41.20 hours. This size is quite significant, given that existing high-to-low-resource S2S pairs have <20hours of parallel audios. Beyond building S2ST models, BENYO-S2ST-Corpus-1 can be used to build pretrained models or improve existing ones for either of the languages(most especially the highly low resourced Yorùbá) towards other downstream tasks such as Text2Speech (TTS), direct Speech2Text(S2T), Automatic Speech Recognition (ASR) and Neural Machine Translation (NMT). Furthermore in this study, we utilised the corpus and Coqui framework to build a pretrained Yorùbá TTS model (named YoruTTS-1.5) as a proof of concept. The YoruTTS-1.5 gave a F0 RMSE value of 63.54 after 1,000 epochs, which indicates moderate fundamental pitch similarity with the reference real-time audio. Ultimately, the corpus architecture in this study can be leveraged by researchers and developers to curate datasets for multilingual high-resource-to-low-resource African languages. This will bridge the huge digital divides in translations among high and low resource language pairs. BENYO-S2ST-Corpus-1 and YoruTTS-1.5 are publicly available on HugginFace at (https://bit.ly/40bGMwi) and (https://bit.ly/3GtUKmH).

## 1.0 Introduction

Speech-to-Speech Translation (S2ST) plays a crucial role in breaking language barriers and fostering effective communication across diverse linguistic communities (Jia et al., 2019). Yorùbá has about 47 million speakerbase and it is one of the official languages in Yorubaland which extends from the southwestern part of Nigeria into Benin Republic and Togo. Other countries where Yorùbá is spoken include Sierra Leone, Ghana, Brazil and Cuba (Ahia et al, 2024). Despite the large speakerbase, Yorùbá (together with other African languages) are endangered due to the dominance of English(or French) as the lingua franca in almost all official conversations in the listed countries. This is clearly reflected in the low resourcefulness of the language in digital domains, thereby hindering inclusive digital tools accessibility by millions of Yorùbá speakers that can not read, write or understand the English language (Goldhahn et al, 2016). Thus, an accurate and efficient English-to-Yorùbá S2ST system is essential for inclusive access to digital tools in education, healthcare, governance as well as socio-economic and religious conversations (Adebola et al., 2020).

Nonetheless, one of the major challenges in developing robust S2ST systems for low resourced languages like Yorùbá is the scarcity of high-quality parallel speech datasets (Anastasopoulos & Chiang, 2018). Building a robust bilingual English-to-Yorùbá S2ST corpus is a complex task that involves curation of high-quality, domain-relevant, and naturally sounding audio. Notably, the conventional approaches for creating such corpus rely on human recordings by expert speakers of the two languages, which are time-consuming, expensive, and prone to inconsistencies (Zoph et al., 2016). Also, corpora created with such a conventional approach are either too small for efficient speech processing or are single speaker and single domain. Furthermore, it requires a large number of fluent bilingual speakers, which could have humongous cost implications(Gutkin et al., 2020a; Jia et al, 2022).

However, recent advances in artificial intelligence, such as data augmentation or synthetic data curation provides promising solutions to the aforelisted challenges (Zenkel et al., 2023). Automated data curation is an approach that leverages state-of-the-art pre-trained or fine-tuned AI based language models to clean, align, and synthesize speech data. This has inherent capability to enhance the efficiency and scalability of corpus creation while maintaining high linguistic fidelity (Wang et al., 2021).

Thus, this work proposes the curation of Bilingual English-to-Yorùbá speech-to-speech translation corpus (BENYO-S2ST-Corpus-1) from both real-time recordings and synthetic audios generated with AI models. This approach is intended to address the limitations of existing datasets for this language pair. This is to further lay a foundation for the development of more accurate and contextually aware speech-to-speech translation models. By employing hybrid data curation, the corpus aims to provide large scale, high-quality, and scalable digital resources for researchers and developers working on English and Yorùbá speech technologies and/or bidirectional English-Yorùbá speech pair. The work also contributes to the growing field of African language technology by introducing an innovative approach to large scale speech corpus curation for low resource languages. This effort would ultimately mitigate digital inequality in line with the Sustainable Development Goal 3 (SDG 3). It will also enhance usability of low resourced languages (i.e. Yorùbá) in AI-driven communication systems for critical domains such as health, education, governance, and commerce.

The rest of this paper is structured as follows: Section 2 presents the background knowledge on Yorùbá language and reviews existing work on language corpora, especially mapping low to high resource language pairs. Section 3 outlines the methodology while Section 4 presents the results and discussion. Finally, Section 5 concludes the study and presents future directions.

## 2.0 Background and Related Works

### 2.1 Yoruba Language Overview

Yorùbá language, commonly spoken by an estimated 47 million people (across countries earlier listed in Section 1.0) is considered a low resource language (CIA, 2025; Gutkin et al. 2020; Ahia et al. 2024). Belonging to Yoruboid sub-category of the Benue-Congo branch of the Niger-Congo family (Hammarström et al. 2019), Yorùbá is the second largest spoken language in Nigeria (Simons & Fennig 2018). Twenty five (25) letters constitute the Standard Yoruba (SY) alphabets with eighteen (18) consonants represented graphemically by *b, d, f, g, gb, h, j, k, l, m, n, p, r, s, ṣ, t,, w, y* and seven (7) vowels by *a, e, ẹ, i, o, ọ, u*. In addition, the language has five (5) nasal vowels represented graphemically as *an, ẹn, in, ọn, un*. It is also made up of five (5) syllable structures of *oral vowels (V)*, *nasal vowels (Vn)*, *syllabic nasals (N)*, combination of *consonants and oral vowels (CV)* as well as combination of *consonants and nasal vowels (CVn)*.

The Standard Yorùbá (SY) language is normally used by the native and other speakers in language education, mass media, and everyday conversation (Ahia et al. 2024). However, it is considered an endangered language due to the dominance of English(or French) in various sectors including education, public health communication, inter-governmental relations, commerce, religious communications as well as social interactions in government parastatals and corporate organisations (Oparinde 2017). The relative paucity of digital resources for Yorùbá language are well noted by researchers and language technologists (Goldhahn 2016). Therefore, this has triggered the interests of researchers in Africa and other parts of the world to evolve innovative strategies towards enhancing the resourcefulness of the language (Adetunmbi 2016; Iyanda 2017; Gutkin et al. 2020).

## 2.2 Related Works

A lot of corpora have been created by researchers for speech translation or transcription across low and high resourced languages. They generally contain texts or audio pairs of the source and target languages. For instance, TEDx Corpus (Salesky et al., 2021) is a dataset containing audio recordings collected during TEDx talks in 8 different source languages. It was created by transcribing the audio files and segmenting them into sentences as well as aligning target language transcripts to the source language audio files. This is to support speech recognition and speech translation tasks for the supported languages. The corpus and corresponding source codes were open-sourced to enable further extensions and improvement within the NLP research community.

Ógúnrèmí et al. (2024) introduced the ÌròyìnSpeech, which is a contemporary Yorùbá speech corpus curated from about 23,000 text sentences from news and creative writing domains. About 5,000 sentences were made available to the Mozilla Common Voice platform (Ardila et al, 2019) to crowd-source human recordings and validation of the Yorùbá speech data by Volunteers. Thus, the corpus contains 6 hours of validated recordings on Mozilla platform and 42 hours of recorded speech from 80 volunteers outside of the platform. The dataset is suitable for Text-to-Speech (TTS) and Automatic Speech Recognition (ASR) tasks. The TTS evaluation indicates the possibility of generating a high-fidelity, general domain and single-speaker Yoruba voice with 5 hours of speech.

The medical sector in Africa is plagued with a very high patient-to-doctor ratio and issues linked to the inability of clinicians to clearly understand their patients (Olatunji et al. 2023). Conversely, a lot of progress has been made in this direction in developed climes through the development of ubiquitous ASR systems. Such technological advances are grossly lacking within the African clinical domains. Where they exist albeit scantily, gaps such as racial biases and minority accents still hinder acceptance for production deployment. Thus, Olatunji et al. (2023) created AfriSpeech-200 to address the lack of accented clinical datasets for building ASR systems towards deployment within the African healthcare systems. AfriSpeech-200 is a publicly available corpus containing 200 hours of Pan-African accented English speech. It contains 67,577 clips from 2,463 unique speakers across 120 indigenous accents from 13 African countries for clinical and general domain ASR. The dataset was used to train ASR systems that achieved state-of-the-art performances.
YORÙLECT(Ahia et al. 2024) is a high-quality parallel text and speech corpus designed from three domains (i.e. news, religion, and TED talk), standard Yorùbá, and four regional Yorùbá dialects (i.e. Ife, Ondo, Ijebu and Ilaje). The corpus was built through extensive fieldwork by the authors interfacing with native speakers in Nigeria's South West geopolitical zone to collect speech recordings that correspond to text transcripts presented to the natives. Several experiments were conducted on text-to-text translation, speech-to-text translation, and automatic speech recognition. The authors reported that the corpus would greatly contribute to the development of NLP models for Yorùbá and its dialects.
CMU Wilderness (Black et al., 2019) is a multilingual speech dataset containing over 700 different languages (including Yorùbá) that provide aligned text, audio, and word pronunciations. For each language in the corpus, there are approximately 20 hours of transcriptions. The corpus was used to design a speech synthesizer using multipass alignment techniques, which was acclaimed by the authors as being good enough for deployment.
Parallel texts are critical resources when performing cross-lingual transfer among low and high resourced languages. JW300 (Agic & Vulic, 2019) is a parallel corpus designed to address the shortage of parallel texts in the NLP domain. The corpus contains 300 languages with 100 thousand parallel sentences per language pair. The utility of the corpus was shown using an experiment with the multi-source part of speech and word embedding induction.

Impressive capabilities have been shown by massively multilingual Machine Translation(MT) systems when performing few and zero-short translations among low-resource languages. Most multilingual models are evaluated on high-resource languages with the assumption that they will generalize on low-resources languages. The lack of standardized evaluation datasets for low-resourced languages makes it difficult to evaluate MT in such languages. The first multi-domain parallel corpus for low-resource pair languages like Yorùbá-English was created by Adelani et al. (2020) and named MENYO 20K. The corpus has a standardized benchmark for train-test splits. Neural MT benchmarks conducted on this dataset outperformed popular pre-trained MT models with a major gain of BLEU +99 and 86. Such benchmark models include Facebook's M2M-100 and Google's multilingual NMT.

## 3.0 Methodology

Synthetic or automatically generated speeches in high-resourced languages such as English are now ubiquitous. They are being applied across different domains such as synthetic content generation, podcasts, voice-over for presentation slides, reading of text for assistive learning, audio feedback in different software solutions and plugins in various applications, which are based on large language models (Ogun et al., 2024). The architecture we developed in this study, (as presented in Figure 1.0) is targeted at curating large scale dataset for direct speech to speech translation ( e.g. English-to-Yorùbá language pair and other language pairs) at reduced cost by leveraging existing non speech-to-speech corporal and pre-trained AI models. The dataset can also be leveraged to build new models for either of the languages to build new models or improve existing ones for other downstream tasks such as TTS, ASR and NMT. The components of the architecture are described in the subsequent subsections.

### 3.1 Data Acquisition

As shown in the first block in Figure 1, this involves downloading existing speech data(with their corresponding transcripts) from various open repositories such as Hugging Face, Kaggle, or GitHub. For this version of the BENYO-S2ST-Corpus-1, we downloaded the YORÙLECT Corpus (Ahia et al. 2024) from the project's Google Drive and extracted the Standard Yorùbá(SY) variant. According to the authors, the dataset is released under an open license, and it can be used in MT(text-to-text), ASR, TTS synthesis, and speech-to-speech translation (S2ST) tasks. Notably, SY is the generic version of the language that has standard orthography. Speakers of the other dialectical variants can understand and speak it regardless of whether they can read or write the transcripts. Also, the majority of published NLP works have been done in SY, and official communication in the language is done with this version. This makes it the most suitable variant for building a speech-to-speech corpus for pairing with English language. Thus, we focused on the SY portion of the corpus, and according to the authors, the SY transcripts in YORÙLECT were obtained from three existing open datasets across three different domains as presented in Table 1.0. As shown, YORÙLECT corpus contains a total of 1,506 SY sentences and the corresponding English transcripts. The version we downloaded for this work contains three metadata files namely *train*, *test* and *validation* with 802, 502 and 200 sentences respectively (N=1,504) for each of SY and English. Comparing this with the total sentence reported in Table 1.0, we suspect that two of the sentences got missing during packaging of the Corpus for open access publishing. Other important attributes in each of the metadata files include, *filename*(*i.e Yoruba audio filename*), *dialect_id, dialect_domain, transcription(SY text), domain, english_text and id.*

**Table 1.0: Sources of Transcripts in YORÙLECT Corpus**

| S/N | Transcript Source | Domain | Number of Sentences | Reference |
|---|---|---|---|---|
| 1 | Bible Study Manual | Religion | 532 | https://faithrebuilder.org/conference-bible-study-manuals |
| 2 | Yorùbá Section of MTTT(i.e. a collection of multi-target bitexts) | TED Talk | 247 | Duh, 2018 |
| 3 | Yorùbá news articles within the MAFT corpus | News | 907 | Alabi et al., 2022 |
| **Total** | | | **1,506** | |

Furthermore, the authors reported that YORÙLECT is the "*first-ever corpus of high quality, contemporary Yorùbá speech and parallel text data across four Yorùbá dialects*". However, exploring the corpus revealed that the English transcripts were rendered in plain text while the equivalent Yorùbá transcripts were garbled and rendered in unreadable non-plain text formats as shown in column 3 of Table 2.0. Further analysis through ChatGPT 4o, which is a state-of-the-art reasoning AI model, inferred the encoding as UTF-8 with 99% confidence (OpenAI, 2025). We suspect that due to Yorùbá having diacritics, as well as special characters like ẹ, ọ, ṣ and *gb*, they were not properly rendered because they were saved with incompatible character set with UTF-8 encoding. Excerpts of the first five raw English-Yorùbá text pairs from the *train* metadata of the YORÙLECT corpus are presented in Table 2.0. Also, as shown in S/N 1, Column 3, the Yorùbá transcript that is readable lacks tonal marks or diacritics, which is critical for deciphering the semantics of the spoken Yorùbá sentence.

**Table 2.0: Excerpts From the Train Metadata of the YORÙLECT Corpus**

| S/N | File-name for Yorùbá audio | Yorùbá Transcript | English Transcript |
|---|---|---|---|
| 1 | data/recorder_2024-01-13_11-24-41_453538.wav | awon apositeli, awon woli, awon ajinrere ati awon oluso agutan ati awon oluko. | Apostles, Prophets, Evangelists and shepherds and teachers. |
| 2 | data/recorder_2023-04-10_11-52-43_936741.wav | G·∫πÃÅg·∫πÃÅ b√≠ Ol√≥y√® √ìg√∫nt√≥s√¨n ti ·π£e r√≤ f√∫n Oh√πn √Ägb√°y√© n√≠n√∫ √¨f·ªçÃÄr·ªçÃÄw√°ni l·∫πÃÅnuw√≤ l√≥r√≠ ·∫πÃÄr·ªç ayeÃÅlujaÃÅra | As Chief Oguntosin told the Voice of the World in an online interview |
| 3 | data/recorder_2023-04-20_13-03-32_758384.wav | IÃÄyawo DaÃÅfiÃÅdiÃÄ ko arun koÃÄroÃÅnaÃÄ eleyi ti ·ªçk·ªç r·∫π kede faye ni ·ªçs·∫π to k·ªçja. | David's wife caught corona, the one her husband was announcing to the world last week |
| 4 | data/recorder_2023-04-10_16-37-52_807071.wav | Aar·∫π nigba naa tun pada dije fun ipo aar·∫π, ·π£ugb·ªçn ko ja m·ªç l·ªçw·ªç. | The president then contested for president again but he did not win |
| 5 | data/recorder_2023-04-20_16-23-50_315100.wav | Ol√≥y√® √ìg√∫nt√≥s√¨n n√≠gb√†gb·ªçÃÅ w√≠ p√© Od√πduw√†, t√≠ √≥ j·∫πÃÅ b√†b√° ≈Ñl√° √¨ran Yor√πb√° lo al√≠f√°b·∫πÃÅ·∫πÃÄ t√¨ n√°√† n√≠ ay√© √†tij·ªçÃÅ. | Chief Oguntosin consistently believed that in ancient times, Oduduwa, the progenitor of the Yoruba race, utilized the alphabet. |

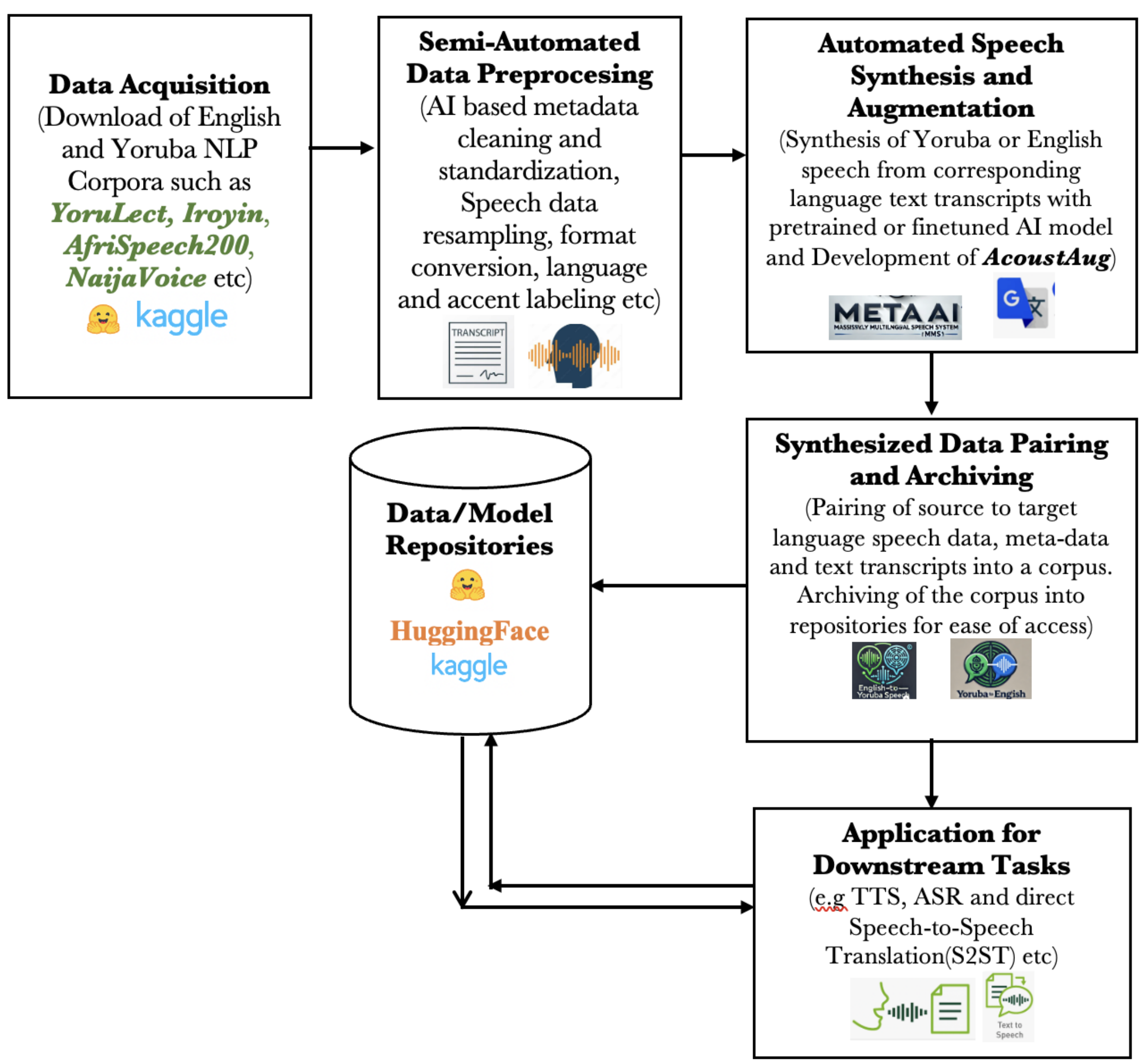

**Figure 1.0: Hybrid(Synthetic and Real-Time) Speech-to-Speech Curation Architecture**

### 3.2 Semi-Automated Data Preprocessing

The semi-automated data preprocessing involves AI based metadata cleaning, standardization, speech data resampling, format conversion, language labeling, etc. This procedure is captured in the second block of Figure 1.0. As earlier established in Section 3.1, the garbled Yorùbá texts were cleaned using a semi-automated approach in this work. We developed the ***ASPMIR-Machine-Translation-Testbed for Low Resourced African Languages*** and deployed it on the ASPMIR HuggingFace Space[1]. As shown in Figure 2, the application contains different pretrained/finetuned text-to-text machine translation models that are openly available on HuggingFace for use by researchers and developers. Exploring available models for English-to-Yorùbá text translation unveiled the *Davlan/m2m-100_418M* model to be of high quality with BLEU Score = 1.0 and focusing on news and general domains. Thus, some members of our team were assigned to meticulously translate the readable English sentences in the YORÙLECT corpus to their Yorùbá equivalent transcripts using *Davlan/m2m-100_418M* model. Because the selected team members are fluent bilingual English and Yorùbá speakers, they ensured that the semantics of the translated texts match their English equivalents. Moreso, the model captured diacritics annotation of the resulting Yorùbá text, which is critical for semantic understanding. The activity took about 7 days to fully translate and ascertain the quality of all the 1,504 sentences in the curated SY YORÙLECT corpus.

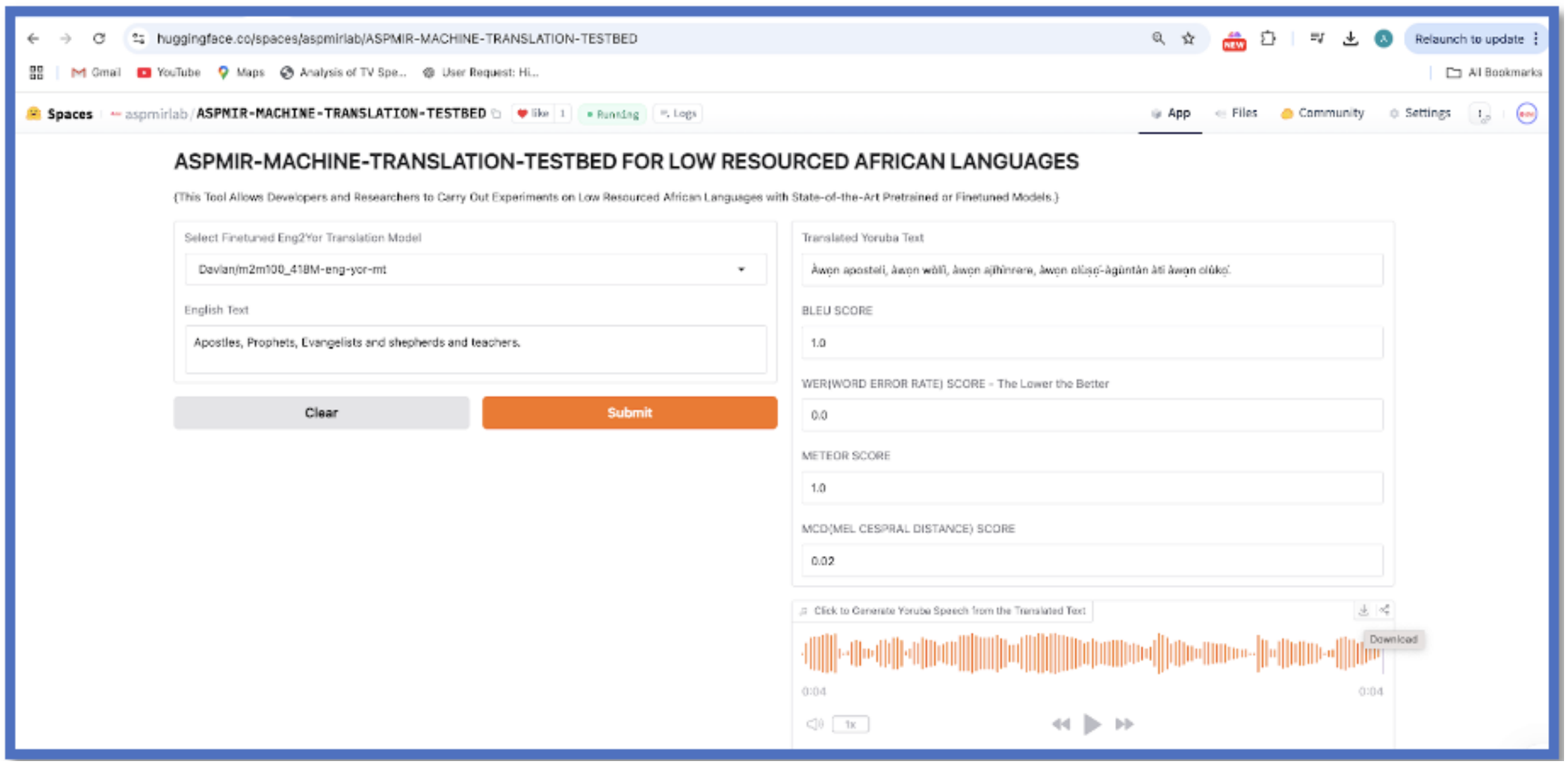

**Figure 2.0: ASPMIR-Machine-Translation-Testbed for Low Resourced African Languages**

[1](https://huggingface.co/spaces/aspmirlab/ASPMIR-MACHINE-TRANSLATION-TESTBE)

Furthermore, the audio files of the YORULECT corpus were analysed using a Python script developed in this study to determine the duration and sampling rates of each file. As shown in Figure 3.0 (with five of the audio files displayed), the duration of the files are expectedly different while the sampling rate is 48kHz. Nonetheless, we developed Python scripts to automate the downsampling of all the audio files (.wav) to 22.05kHz.

File: recorder_2023-04-10_17-24-55_196322.wav, Duration: 7.71 sec, SamplingRate: 48.00 kHz

File: recorder_2023-04-10_11-21-06_265115.wav, Duration: 10.09 sec, SamplingRate: 48.00 kHz

File: recorder_2023-04-10_17-14-32_859292.wav, Duration: 11.10 sec, SamplingRate: 48.00 kHz

File: recorder_2023-04-20_14-44-47_889858.wav, Duration: 8.47 sec, SamplingRate: 48.00 kHz

File: recorder_2024-01-13_15-44-08_406739.wav, Duration: 7.83 sec, SamplingRate: 48.00 kHz

**Figure 3.0: Characteristics of Selected Yorùbá Audio Files in the YORÙLECT Corpus Before Downsampling to 22.05kHZ.**

### 3.3 Automated Speech Synthesis and Augmentation

Automated speech synthesis and augmentation is a major component of the architecture as shown in Figure 1.0. This block handles synthesis of either English or Yorùbá transcripts to the speech equivalent using open pretrained or finetuned model as clearly indicated in the block. This is similar to the procedure in (Li et al., 2025) for TTS based synthetic data generation and augmentation for low resource languages such as *Bemba*, *North Levantine Arabic*, and *Tunisian Arabic*. Since the YORÙLECT corpus already contains real-time curated speech in Yorùbá, we leveraged the META AI's Massive Multilingual System (MMS)(Pratap et al., 2024) to automatically synthesize the single speaker audio equivalents of the cleaned 1,504 standard English sentences in the corpus. The custom built application presented in Figure 5.0 was utilised by members of the team for this task, and the resulting audio files were saved into a dedicated Google Drive folder. The duration of each of the synthesized audio(.wav) files are different while the sampling rate is 16kHz (See Figure 4.0). Furthermore, we utilised Python scripts developed in this work to upsample each of the audio files to generate 22.05kHz equivalent audios.

```
File: recorder_2024-01-13_12-28-58_029504.wav, Duration: 4.61 sec, SamplingRate: 16.00 kHz

File: recorder_2024-01-13_12-22-08_507538.wav, Duration: 7.60 sec, SamplingRate: 16.00 kHz

File: recorder_2023-04-10_10-58-10_577619.wav, Duration: 4.46 sec, SamplingRate: 16.00 kHz

File: recorder_2024-01-13_11-26-04_710718.wav, Duration: 3.70 sec, SamplingRate: 16.00 kHz

File: recorder_2023-04-10_16-17-33_551871.wav, Duration: 6.10 sec, SamplingRate: 16.00 kHz
```

**Figure 4.0: Characteristics of Selected Synthesized English Audio Files Based on the English Transcripts in the YORÙLECT Corpus**

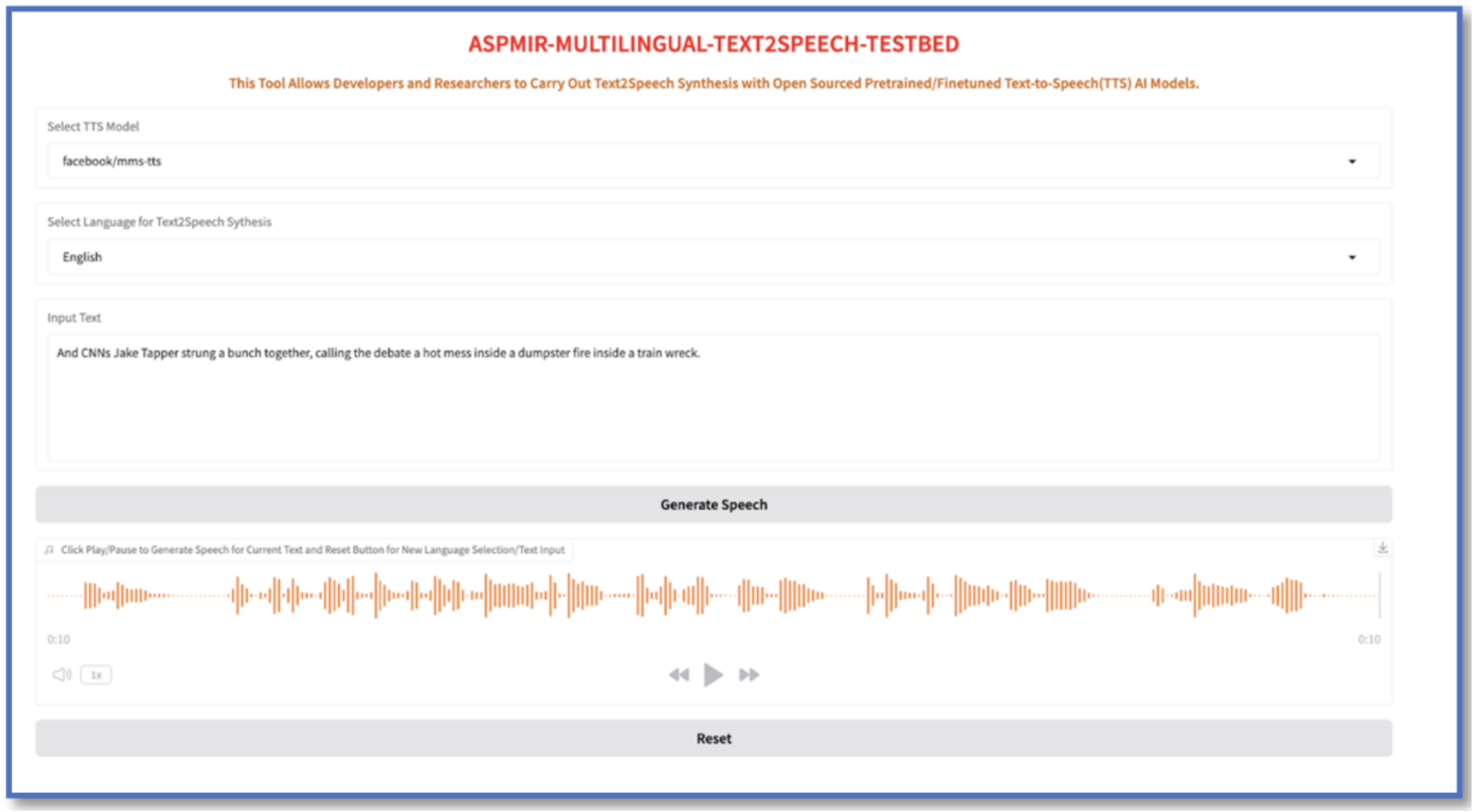

**Figure 5.0: ASPMIR Multilingual Text2Speech Testbed**

Inspired by prior works on audio augmentation (Li et al., 2025; Robinson et al., 2022; Park et al., 2019), we developed **AcoustAug,** a new audio augmentation algorithm to generate augmented audios from both synthetic and raw audio files for English and Yorùbá languages respectively. The algorithm is based on three key acoustic and latent audio variables, namely: *i) speed (0.9 and 1.1 factors)*, *ii) pitch (0.95, 1.05 factors)* and *iii) volume (-5dB, +5dB, +10dB factors)*. This increased the number of English and Yorùbá audio files and their corresponding replicated (i.e. oversampled) transcripts to eight folds each (i.e. 1,504 x 8 = 12,032) with a total of 24,064 samples for the two languages. The full procedure for **AcoustAug** is presented in Algorithm 1. The algorithm and other codes in this study were implemented with Python programming language on our private cloud infrastructure for collaborative programming named CodingHub (https://codinghub.fedgen.net).

**Algorithm 1: AcoustAug - An Audio Data Augmentation Algorithm**

**INPUT:** input_folder, output_folder, aug_type(either “speed”,”pitch”,”volume”)

**OUTPUT:** augmented_audio_files(.wav files with applied speed, pitch or volume saved in output_folder

```
FUNCTION AugmentAudio(input_folder, output_folder, aug_type, sample_rate):
   // Ensure the output directory exists
   CREATE output_folder IF IT DOES NOT EXIST
   // Iterate through each file in the specified input folder
   FOR EACH file IN input_folder:
      // Process only WAV files
      IF file HAS '.wav' EXTENSION THEN
         SET file_path = FULL PATH TO current file
         SPLIT file name INTO name AND extension (e.g., "audiofile", ".wav")
         // Load the audio and resample it for processing
         LOAD audio FROM file_path WITH initial sample_rate = 48000 Hz
         RESAMPLE loaded_audio TO 22050 Hz AND ASSIGN TO audio_22k_raw

         // Perform augmentation based on the specified type
         IF aug_type IS "speed":
            // Apply speed changes
            FOR EACH factor IN [0.9, 1.1]: // Iterate through speed factors (90% and 110%)
               CREATE RESAMPLE EFFECT WITH new_rate = sample_rate * factor
               CONVERT audio_22k_raw TO TENSOR (waveform_tensor)
               APPLY speed effect TO waveform_tensor AND STORE IN augWaveform
               SET output_path = output_folder + name + "_22k_" + factor + "_speed" + extension
               WRITE augWaveform TO output_path WITH sample_rate 22050 Hz
               PRINT "Processing speed augmentation for " + name + " with factor " + factor
           END FOR
         END IF

         ELSE IF aug_type IS "pitch" THEN
            // Apply pitch changes
            CONVERT audio_22k_raw TO TENSOR (waveform_tensor)
            FOR EACH pitch_factor IN [0.95, 1.05]: // Iterate through pitch factors (95% and 105%)
               CREATE UPSAMPLE EFFECT TO rate = sample_rate * pitch_factor
```

```
          APPLY upsample THEN downsample TO waveform_tensor AND STORE IN pitchAugWaveform
          SET output_path = output_folder + name + "_22k_" + pitch_factor + "_pitch" + extension
          WRITE pitchAugWaveform TO output_path WITH sample_rate 22050 Hz
          PRINT "Processing pitch augmentation for " + name + " with factor " + pitch_factor
        END FOR
      END IF

      ELSE IF aug_type IS "volume" THEN
        // Apply volume changes
        CONVERT audio_22k_raw TO TENSOR (waveform_tensor)
        FOR EACH dB_change IN [-5, 5, 10]: // Iterate through dB changes (-5dB, +5dB, +10dB)
          // Adjust volume using a decibel-to-amplitude conversion formula
          SET volAugWaveform = waveform_tensor * 10^(dB_change / 20)
          SET output_path = output_folder + name + "_22k_" + dB_change + "_vol" + extension
          WRITE volAugWaveform TO output_path WITH sample_rate 22050 Hz
          PRINT "Processing volume augmentation for " + name + " with dB change " + dB_change
        END FOR
      END IF
END FUNCTION
```

### 3.4 Data Pairing and Archiving

Pairing of audio files across languages (with their corresponding transcripts) as well as systematic archiving play a crucial role in the development of (multi/bi)lingual speech technologies (Li et al., 2025). This is especially critical for tasks such as speech translation as well as ASR and TTS. We ensured sample rate alignment (22.05kHz) for all the augmented English and Yorùbá audio files. For proper file level alignment, each of the folders contain audio files of the source and target languages respectively (see Figure 6.0). Notably, the filenames used in the SY component of the YORÙLECT Corpus were maintained for the augmented Yorùbá audios as well as for the corresponding augmented English audios in the current version of the corpus. Furthermore, we generated a metadata (.csv) file with four columns, (i.e. *audio-filename-eng*, *transcript-eng*, *audio-filename-yor*, *transcript-yor*). The metadata is a vital resource that can be leveraged for programmatic alignment of audio pairs for direct S2ST tasks. This is also applicable to text-to-audio and audio-to-text pairs for TTS and ASR tasks respectively for either the source or target language. Although the current edition of the proposed BENYO-S2ST-Corpus-1 is single speaker for each of the source and target languages, it is well structured and archived for reproducibility and scalability.

All the audio folders (i.e. augmented-audio-eng-12k for English and augmented-audio-yor-12k for Yorùbá) as well as the metadata are archived on the Advanced Signal Processing and Machine Intelligence Research (ASPMIR) public repository on HuggingFace(https://bit.ly/40bGMwi) for open access.

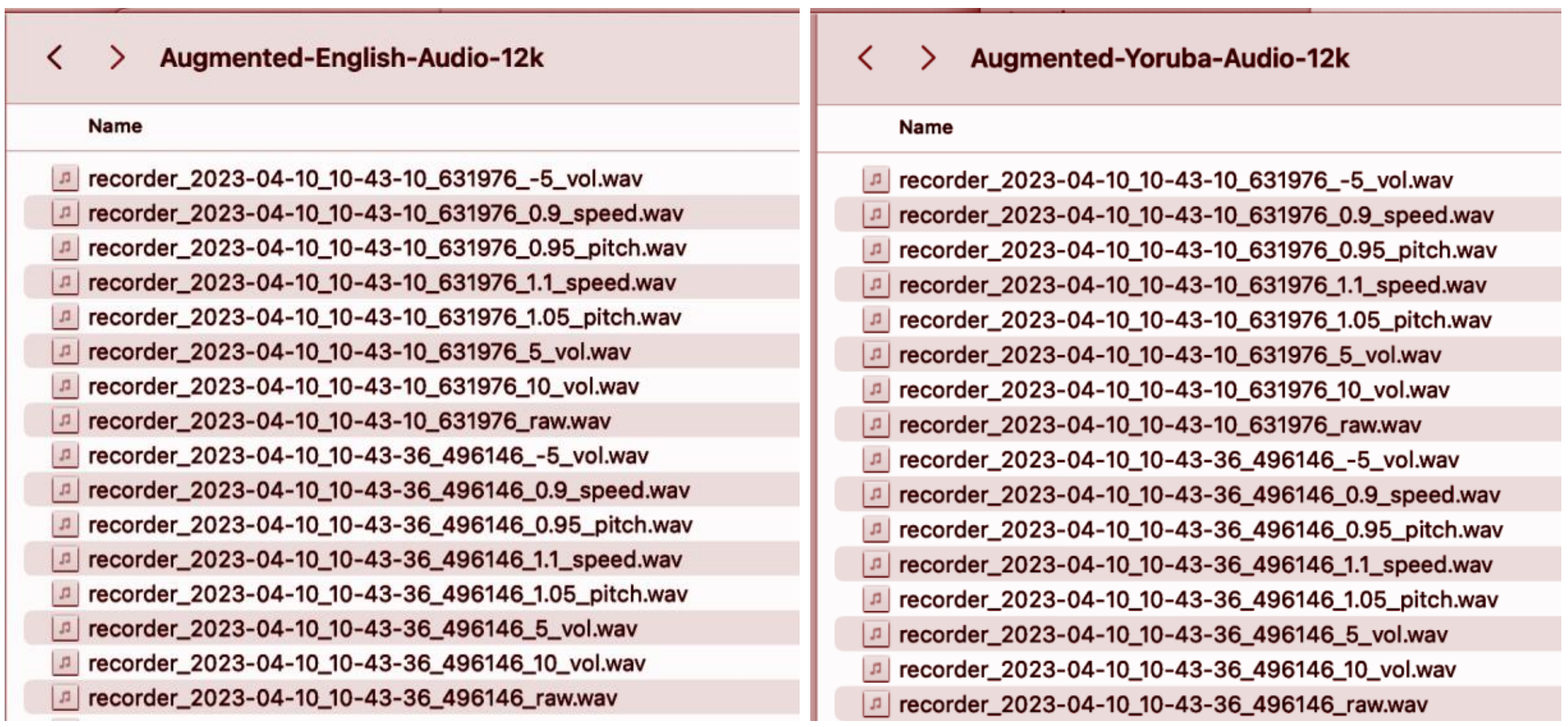

**Figure 6.0: Sample Files from the Augmented English and Yorùbá Audios**

### 3.5 Application of the Corpus for Yorùbá Text-to-Speech Synthesis

We selected Yorùbá TTS task to explore the potentials of the proposed corpus for downstream applications. Notably, using TTS-based augmentation to generate large-scale synthetic English audio is well reported in the literature (Li et al., 2025; Moslem 2024; Robinson et al., 2022) due to its high resourcefulness. Conversely, this is not the case with the Yorùbá Language, which is extremely low resourced in terms of audio datasets and speech based models. The eighteen (18) state-of-the-art TTS models presented in the Appendix across five (5) architectural categories (i.e. *autoregressive*, *flow-based*, *diffusion-based*, *parallel feedforward,* and *prompt-based*) reveal the predominance of the English language. Only META AI's MMS supports the Yorùbá TTS language in its pre-trained mode but with no Yorùbá specific Grapheme2Phoneme(G2P) tool (Pratap et al., 2024). Furthermore, Variational Inference Text-to-Speech(VITS) was not pretrained with Yorùbá but can be finetuned for it. The critical place of robust TTS models for synthesizing target audio files for cascaded and direct S2ST models from a source to target language was also highlighted by Jia et al. (2022) in building the Translatotron 2 model. Thus, given the foregoing, developing a Yorùbá TTS model with the augmented Yorùbá audio and transcript pairs, which is a subset of the proposed BENYO-S2ST-Corpus-1 presents several potential benefits. The major benefit is that the model can be utilised to carry out TTS-based augmentation, which would boost the size of the audio samples for further upgrade of BENYO-S2ST-Corpus-1.

Thus, we developed a new Yorùbá TTS model named **YoruTTS-1.5***,* based on BENYO-S2ST-Corpus-1 using the Coqui TTS framework (Coqui AI, 2025). The framework is an open-source

Python library designed for TTS model pretraining or finetuning. It is modular and extensible with support for autoregressive, non-autoregressive and flow-based TTS models. One of Coqui's most powerful tools, which combines acoustics modeling and vocoding is the Variational Inference Text-to-Speech (VITS). Even though the original version does not cater for the Yorùbá language, the architecture can be adapted and pretrained with Yorùbá audio and text pairs or any other low resource languages (Coqui AI, 2025). Thus, Figure 7.0 shows the architecture of **YoruTTS-1.5** we developed by adapting VITS architecture for Yorùbá TTS, even though it was natively built for English. The components of the architecture are i) Text Input (i.e. Yorùbá transcripts), ii) Text Processing, iii) Acoustic Modeling and Vocoder, iv) Audio Output (synthesized Yorùbá audio).

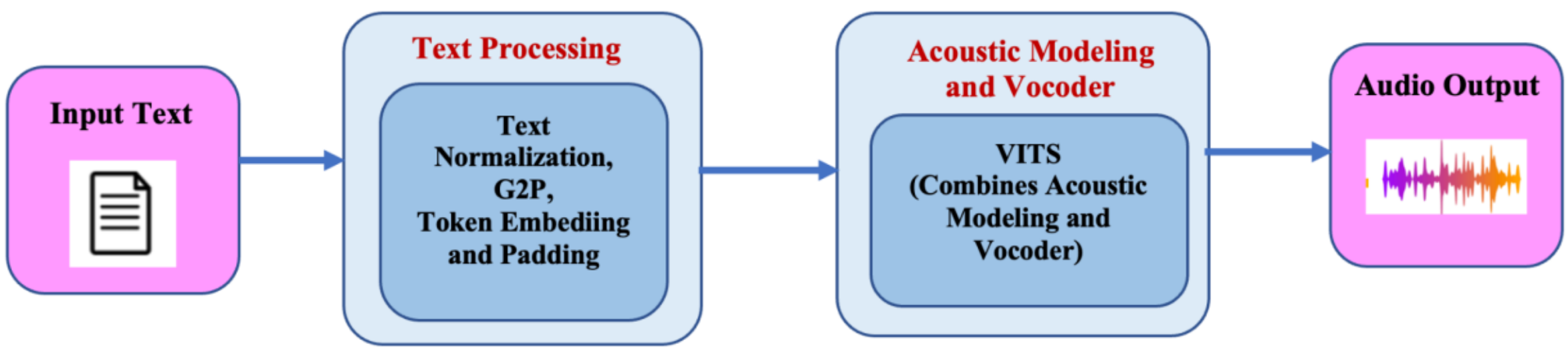

**Figure 7.0: Architecture of YorùTTS-1.5 Model based on Coqui Framework and BENYO-S2ST-Corpus-1**

### 3.5.1 Text Processing

This is the second block in the architecture(Figure 7.0), which transforms input text into a sequence of phonetic tokens (phonemes) that serve as input to the acoustic modeling and vocoder block. The key components of the text processing block are text normalization, Grapheme-to-Phoneme (G2P) conversion, phoneme/token embedding and padding. There are three classes in the Coqui framework for handling text normalisation. These include *BaseDatasetConfig*, *CharactersConfig* and *TTSTokenizer.*

The BaseDatasetConfig class is used to configure how the dataset is structured and interpreted during pre-training or fine-tuning. The configuration includes; the *path* to the dataset, specific *metadata file* to use (in CSV format) and the *formatter* to use for parsing the dataset. Coqui utilises the "ljspeech" predefined formatter, in which the metadata file is parsed to follow the LJSpeech-style, which is: *filename|transcription.* This implies that each line of the formatted

metadata has the audio filename (without extension) and its corresponding text transcription that are separated by a pipe (|). For the Yorùbá language in this work, the transcription utilises UTF-8 encoding and maintains the diacritics, which is critical for the semantics of the text during audio synthesis.

Coqui uses the CharactersConfig class to handle training configurations such as: i) building of the vocabulary of allowed characters/phonemes, ii) embedding the transcripts into integer token IDs, iii) determination of the G2P rules for specific languages and iv) handling of token-level special symbols such as <pad>, <bos>, <eos>. Rather than using the default settings in Coqui, we manually configured CharacterConfig class attributes to address the specific characters in the Yorùbá language as well as its diacritics. Table 4.0 shows the configurations of our CharacterConfig class attributes.

**Table 4.0: Manual Configuration of CharactersConfig Class**

| S/N | Class Attributes | Value |
|---|---|---|
| 1 | Characters | "bdfggbhjklmnprssṣtwyàaáèeéẹ̀ẹẹ́ìiíòoóọ̀ọọ́ùuú<br>BDFGGBHJKLMNPRṢTWYÀAÁÈEÉẸ̀ẸẸ́ÌIÍ<br>ÒOÓỌ̀ỌỌ́ÙUÚ0123456789” |
| 2 | Punctuations | ".,!?;:\()-[]" |
| 3 | Pad | "<PAD>" |
| 4 | EOS | "<EOS>" |
| 5 | BOS | "<BOS>" |
| 6 | BLANK | "<BLK>" |

**Attributes for Yorùbá TTS**

The TTSTokenizer class is a core utility in Coqui for text embedding. It acts as the bridge between the textual and numerical inputs for the model by converting raw text (i.e. phonemes) into token IDs, which can be fed into the subsequent block. Some additional functionalities of TTSTokenizer are; i) reconstruction of token strings from ID list, ii) setting of a custom phoneme/character vocabulary, iv) storage and reloading of vocabulary files (e.g., phonemes.json), which can be reused during inference to ensure consistent token mapping.

### 3.5.2 Acoustic Modelling and Vocoder

Coqui uses the Variational Inference for Text-to-Speech(VITS), which integrates acoustic modeling and vocoder(HiFi-GAN-style decoder). The VITS model (Kim et al., 2021) was developed based on an end-to-end generative sequence modeling architecture for synthesis of high-quality speech. It leverages conditional Variational Autoencoders (VAE) in combination with normalizing flows to model complex speech distributions. The architecture consists of three principal components, namely: *posterior encoder*, *prior encoder*, and *waveform generator*, which models the following conditional distributions:

***i) Posterior distribution*** $\boldsymbol{q_{\Phi}(z|x)}$ over latent variables *z,* given the observed speech input *x*;

***ii) Prior distribution*** $\boldsymbol{p_{\theta}(z|c)}$ over latent variables *z,* conditioned on linguistic features *c* (e.g. phonemes)*;* and

***iii) Data distribution*** $\boldsymbol{p_{\Psi}(y|z)}$**,** or the likelihood of generating waveform *y* from the latent representation *z.*

The posterior encoder parameter ($\Phi$) captures information from the input speech, while the prior encoder parameter ($\theta$) is trained to approximate the prior distribution based on textual input. Notably, the prior distribution, parameterized with $\theta$ outputs a Gaussian distribution ($\aleph(O, I)$) based on the text input *c.* The modeling of the prior is further improved via a *normalizing flow* $\boldsymbol{f}$, which is a sequence of *invertible* and *differentiable transformations* that maps the simple Gaussian distribution (i.e. $\aleph(O, I)$) to a more complex distribution, thereby enabling more expressive mappings in the latent space. Furthermore, the HiFi-GAN-style decoder (Kong et al., 2020) parameters (ψ) are re-trained to generate waveforms of realistic speech synthesis, which is conditioned on the latent variable *z*. To achieve this, the parameters are trained by maximising the conditional log-likelihood of the data *log p(x|c)* through the Evidence Lower Bound (ELBO):

$$log\, p(x|c) \geq E[log p_{\Psi}(\boldsymbol{x}|\boldsymbol{z})] - DKL(q_{\Phi}(\boldsymbol{z}|\boldsymbol{x})||\boldsymbol{p_{\theta}}(\boldsymbol{z}|\boldsymbol{c})) \quad (1)$$

This ELBO objective function aligns the posterior with the prior distributions and ensures accurate reconstruction of the speech waveform *x* from the latent representation *z*. In the Coqui framework, BaseAudioConfig class defines the configuration of various audio feature extraction procedures such as mel-spectrograms, sampling rate, and pre-emphasis. Table 5.0 contains detailed configurations of the BaseAudioConfig class attributes for this work.

**Table 5.0: Configurations of the BaseAudioConfig Class Attributes for YorùTTS-1.5**

| S/N | Class Attributes | Full Meaning | Value |
|---|---|---|---|
| 1 | sample_rate | Sample Rate | 22,050Hz |
| 2 | win_length | Window Length | 1,024 samples |
| 3 | hop_length | Hop Length | 256 samples |
| 4 | num_mels | Number of Mel Filterbanks | 80 |
| 5 | mel_fmin | Minimum Frequency for Mel Filterbank | 0Hz |
| 6 | mel_fmax | Maximum Frequency for Mel Filterbank | None |
| 7 | fft_size | Fast Fourier Transform Size | 1,024 |

### Model

Furthermore, the VITSConfig class is a model-specific configuration class in Coqui that stores all the parameters required to train and evaluate the **YorùTTS-1.5** model. The attributes' specifications for VITSConfig class in developing the pretrained **YorùTTS-1.5** model are presented in Table 6.0.

**Table 6.0: Configuration of VITSConfig Class Attributes for Initialising the Training of YorùTTS-1.5 Model**

| S/N | Class Attributes | Full Meaning | Value |
|---|---|---|---|
| 1 | batch_size | Number of training samples per batch. | 16 |
| 2 | eval_batch_size | Number of validation samples per batch. | 8 |
| 3 | num_loader_workers | Controls how many parallel CPU workers are used to load and pre-process data during training. | 4 |
| 4 | num_eval_loader_workers | Number of parallel data loading workers used during evaluation. | 4 |

| 5 | run_eval | A Boolean flag to enable or disable evaluation during training. | True |
|---|---|---|---|
| 6 | test_delay_epochs | Number of epochs to wait before running the first evaluation (validation/test) during training. | -1 |
| 7 | epochs | Total number of training epochs | 1,000 |
| 8 | text_cleaner | Specifies the text preprocessing function(s) applied to input text before tokenization. | None |
| 9 | use_phonemes | Whether to convert input text to phonemes using a grapheme-to-phoneme (G2P) tool. | False |
| 10 | phoneme_language | Language code used to select the correct G2P rules when use_phonemes=true | None |
| 11 | phoneme_cache_path | Path to a file where preprocessed phoneme sequences are cached, Speeds up training by avoiding repeated phoneme conversion using G2P for every epoch | os.path.join(output_path, "phoneme_cache") |
| 12 | print_step | Frequency (in steps) for printing training logs to the console | 25 |
| 13 | print_eval | Whether to print evaluation results (e.g., validation loss, metrics) after each evaluation cycle. | False |
| 14 | mixed_precision | Enables automatic mixed-precision training (using float16 and float32) for speed and efficiency. | True |
| 15 | output_path | Root directory where all training outputs and artefacts are saved. | output_path (contents include, *model checkpoints*, *logs (train.log)*, *synthesized audio* |
| | | | *samples*, *configuration snapshots*, *tokenizer* and *phoneme cache files*). |
| 16 | datasets | Configuration section that defines the datasets used for training, validation, and testing. | dataset_config (an instance of the BaseDatasetConfig class) |
| 17 | characters | Defines how text is tokenized into characters or phonemes. | yoruba_characters (an instance of CharactersConfig class) |
| 18 | audio | Defines how raw audio files are converted into mel-spectrograms and how they are normalized. | audio_config (an instance of BaseAudioConfig) |
| 19 | use_language_embedding | Flag that enables language conditioning in multilingual TTS models. It adds a learned embedding vector for each language for conditioning the model when synthesizing speech. | False |

The AudioProcessor class is also a core utility in the Coqui framework, which handles all audio-related preprocessing tasks such as loading/saving of audio waveforms, conversion of waveforms' sample rate, normalization of volume based on amplitude or decibel range, conversion of the waveforms to mel(linear)-spectrograms, silence trimming, generation of mel-spectrograms from waveforms, approximation of waveform reconstruction from mel, saving of output waveform during inference, and emphasizing or de-emphasizing of high-frequency contents. It was configured in this work based on the parameters in BaseAudioConfig class (Table 5.0).

## 4.0 Results and Discussion

### 4.1 Statistics of the Augmented Audios

The statistical characteristics of both the standard English and standard Yorùbá augmented audios in the BENYO-S2ST-Corpus-1 are presented in Table 7.0. English and Yorùbá language have 12,032 pairs of audio files per language, which gives a total of 24,064 sample size. The minimum, maximum, average and total durations for English audios are 1.12s, 14.85s, 5.33s and 64,131.71s (17.81h) respectively whereas the values for Yorùbá audios are 1.13s, 16.62s, 7.00s and 84,201.80(23.39h). Thus, the total audio duration for the two languages is 41.20hours. This size is quite significant, given that existing high-to-low-resource S2S pairs have <20h of parallel audios. For instance, the total duration for Yorùbá audios in FLEURS is approximately 15h and it is less than 10h in Common Voice (Conneau et al. 2023; Jia et al. 2022). Furthermore, the International Conference on Spoken Language Translation (IWSLT) 2023 S2ST Challenge focused on real S2S translation for low-resource target languages. The released audio pairs from the challenge for English-to-low resource target languages contain 17h of English–Farsi, 8h of English–Indonesian, 3h of English–Catalan and 4h of English–Vietnamese (Salesky et al., 2023).

**Table 7.0: Augmented Audios Characteristics**

| **Augmented Synthesized Standard English Duration (seconds)** | | | | **Augmented Standard Yorùbá Duration (seconds)** | | | |
|---|---|---|---|---|---|---|---|
| Min. | Max. | Ave. | Total | Min. | Max. | Ave. | Total |
| 1.12 | 14.85 | 5.33 | 64,131.71 (17.81hrs) | 1.13 | 16.62 | 7.00 | 84,201.80 (23.39hr) |
| Total English Audio Samples =12,032 | | | | Total Yorùbá Audio Samples = 12,032 | | | |

### 4.2 Acoustic Analysis of the Original and Augmented Audios

The waveforms of the augmented Yorùbá audios vis-a-vis the original recorded ones (for transcript - *A gbà wá là, nítorí rẹ̀, ní ìrẹ̀tí ìtẹ̀síwájú rẹ̀*) are presented in Figure 8.0. Similar waveforms for the equivalent English transcript (*We are saved by his grace, in the purpose of his progress*) are presented in Figure 9.0. Visual inspection shows that the augmentations across all the latent acoustic features and the corresponding factors (i.e. volume (factors -5, 5 and 10); speed (factors 0.9 and 1.1); and pitch (factors 0.95 and 1.05) have similar wave shapes. This connotes that the ***AcoustAug*** augmentation algorithm preserved the core temporal and spectral structures of the audios for both languages while introducing sufficient variations needed for model's generalisation.

Furthermore, we quantified the acoustic similarity between the original recorded/synthetic audios and the augmented variants for both languages. This was done by using metrics such as Short-Time Objective Intelligibility (STOI), Perceptual Evaluation of Speech Quality (PESQ) and Log-Spectrogram L1 Distance Function (LogSpec-L1). The range of STOI (a measure of audio intelligibility) is from 0 to 1, where 0 represents unintelligibility while 1 stands for perfect intelligibility. The PESQ metric quantifies the perceptual quality of audios with 1 indicating bad quality while >=4.5 represents transparent quality. LogSpec-L1 measures the perceived spectral similarity between a reference and augmented audio with 0-5 value indicating high similarity, 6-10 denoting moderate variation while >10 indicates clear spectral difference.

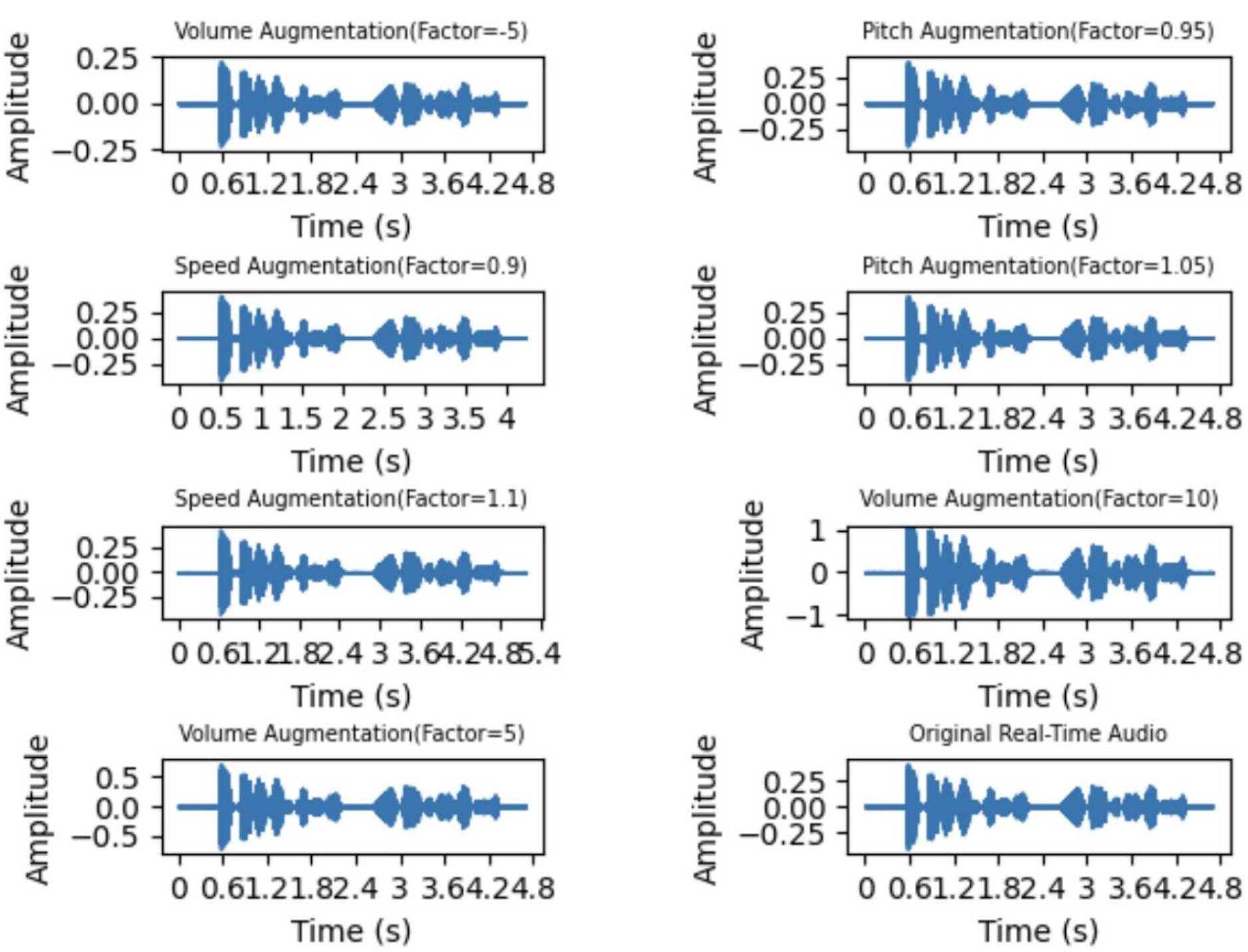

**Figure 8.0: Waveforms of Real and Augmented Yorùbá Audios**
(*Yorùbá Transcript*: A gbà wá là, nítorí rẹ̀, ní ìrẹ̀tí ìtẹ̀síwájú rẹ̀)

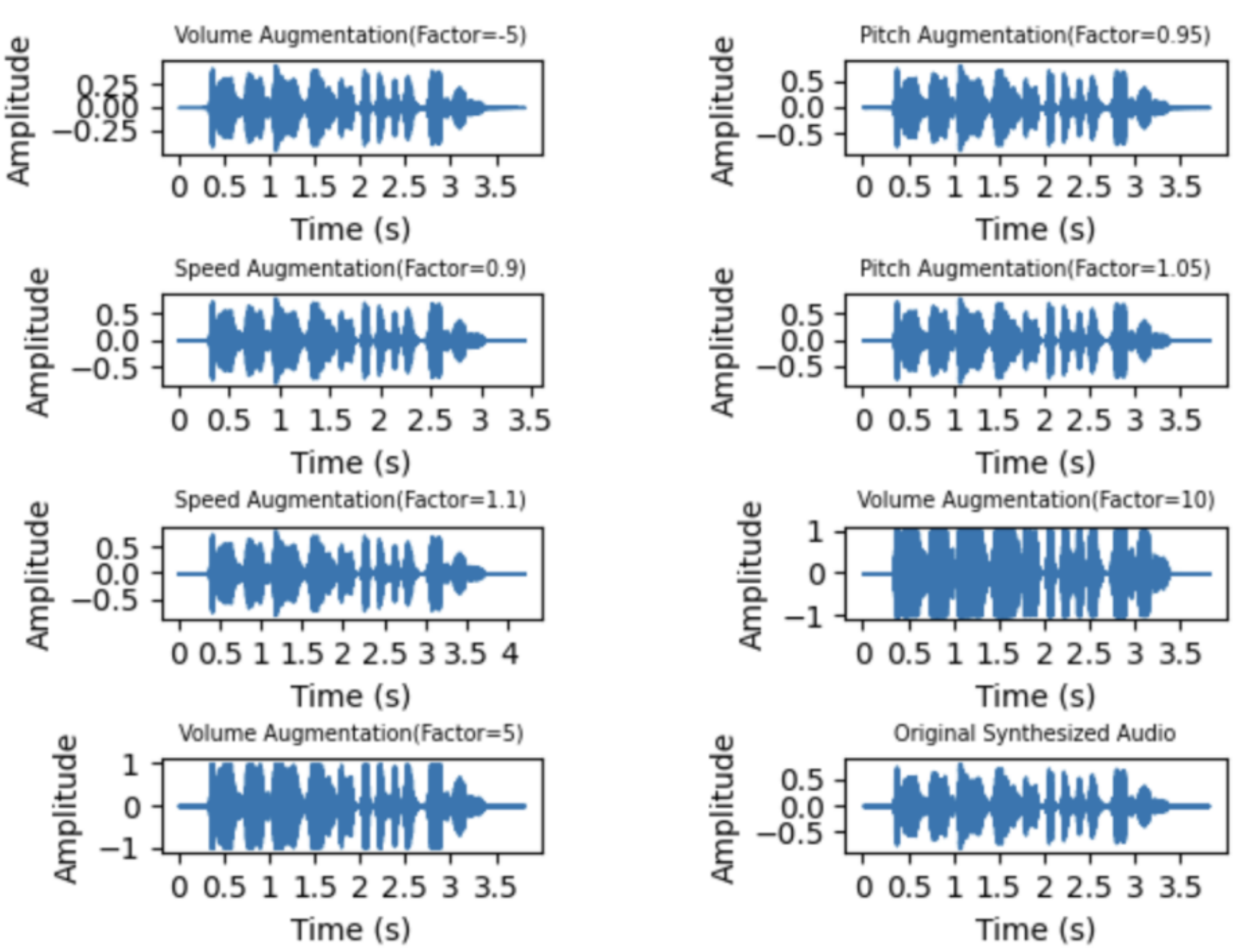

**Figure 9.0: Waveforms of Synthesized and Augmented English Audios**
(*English Transcript*: We are saved by his grace, in the purpose of his progress.
*Domain:* Religion)

Table 8.0 presents the results obtained when a raw audio sample(reference) for each of English and Yorùbá was spectrally compared with the respective augmented versions using STOI, PESQ and LogSpec-L1. Expectedly, computing similarity of the reference audio with itself for each of the languages posted STOI of 1.0, PESQ of 4.6 and LogSpec-L1 of 0.0. The analysis also shows that the augmentation based on volume (factor = -5) and pitch (factors = 0.95, 1.05) posted values that indicate high similarity across all the metrics and for both languages (pink annotation in Table 8.0). This implies that these latent variables appreciably retained the volume and pitch attributes of the reference audio at the indicated factors. It can also be observed that augmentation based on volume (factor = 5) presented higher similarity with the reference audio for Yorùbá language(similar to factor = -5) than for English across all the metrics. Volume augmentation (factor = 10) posted even lower similarity for the two languages (light blue annotation in Table 8). Furthermore, speed based augmentation (factors = 0.9,1.1) presented the lowest similarity for the two languages and across all the metrics. However, based on subjective evaluation by humans, the perceptual qualities and intelligibility of the speed based augmented audios for the two languages are adequate for different use cases notwithstanding the shown quantitative result. Nonetheless, blending augmented audio files of varying similarities with the reference audio in the BENYO-S2ST-Corpus-1 is a critical strategy for introducing sufficient sample variations. This possesses potential benefits in developing models (such as TTS, ASR, S2T and S2ST) with acceptable generalisation capabilities.

**Table 8.0: Analysis of the Original and Augmented Audios**

| Original/Augmented Audio ID | STOI (Intelligibilty) | | PESQ (Perceptual Quality) | | LogSpec-L1 (Perceived Spectral Similarity) | |
|---|---|---|---|---|---|---|
| | English | Yorùbá | English | Yorùbá | English | Yorùbá |
| audio1_reference | 1.0000 | 1.0000 | 4.6440 | 4.6440 | 0.0000 | 0.0000 |
| audio1_vol_-5 | 1.0000 | 1.0000 | 4.6420 | 4.6440 | 2.4470 | 0.0220 |
| audio1_vol_5 | 0.9904 | 1.0000 | 1.1200 | 4.6430 | 2.7430 | 0.0120 |
| audio1_vol_10 | 0.9487 | 0.9999 | 1.0370 | 2.9960 | 4.8460 | 0.0310 |
| audio1_pitch_0.95 | 1.0000 | 1.0000 | 4.6420 | 4.6440 | 2.2380 | 0.1920 |
| audio1_pitch_1.05 | 1.0000 | 1.0000 | 4.6420 | 4.6440 | 2.2110 | 0.1230 |
| audio1_speed_0.9 | 0.0469 | 0.1083 | 1.0650 | 1.1390 | 4.9180 | 3.2950 |
| audio1_speed_1.1 | 0.0357 | 0.1208 | 1.0970 | 1.1010 | 4.5940 | 3.3140 |

### 4.3 TTS Model Training Results

Figure 10.0 shows the YoruTTS-1.5 training performance plots with the various evaluation metrics that were used during the model training. As shown in the plot, the audio quality improved gradually during training with the generator attaining stability (as indicated through the Avg_loss_gen*(*generator loss*)* and Avg_loss mel*(*mel spectrogram reconstruction loss*))* plots. It can also be observed through the Avg_loss_duration(duration prediction loss) that the time it takes for the model to convert text into speech dropped drastically during the early stage of the evaluation step and maintained relatively stable values afterward. Remarkably, the slight increase and then stability of the Avg_loss kl(Kullback-Leibler divergence) and Avg_loss_feat (feature matching loss) are due to regularization during training.

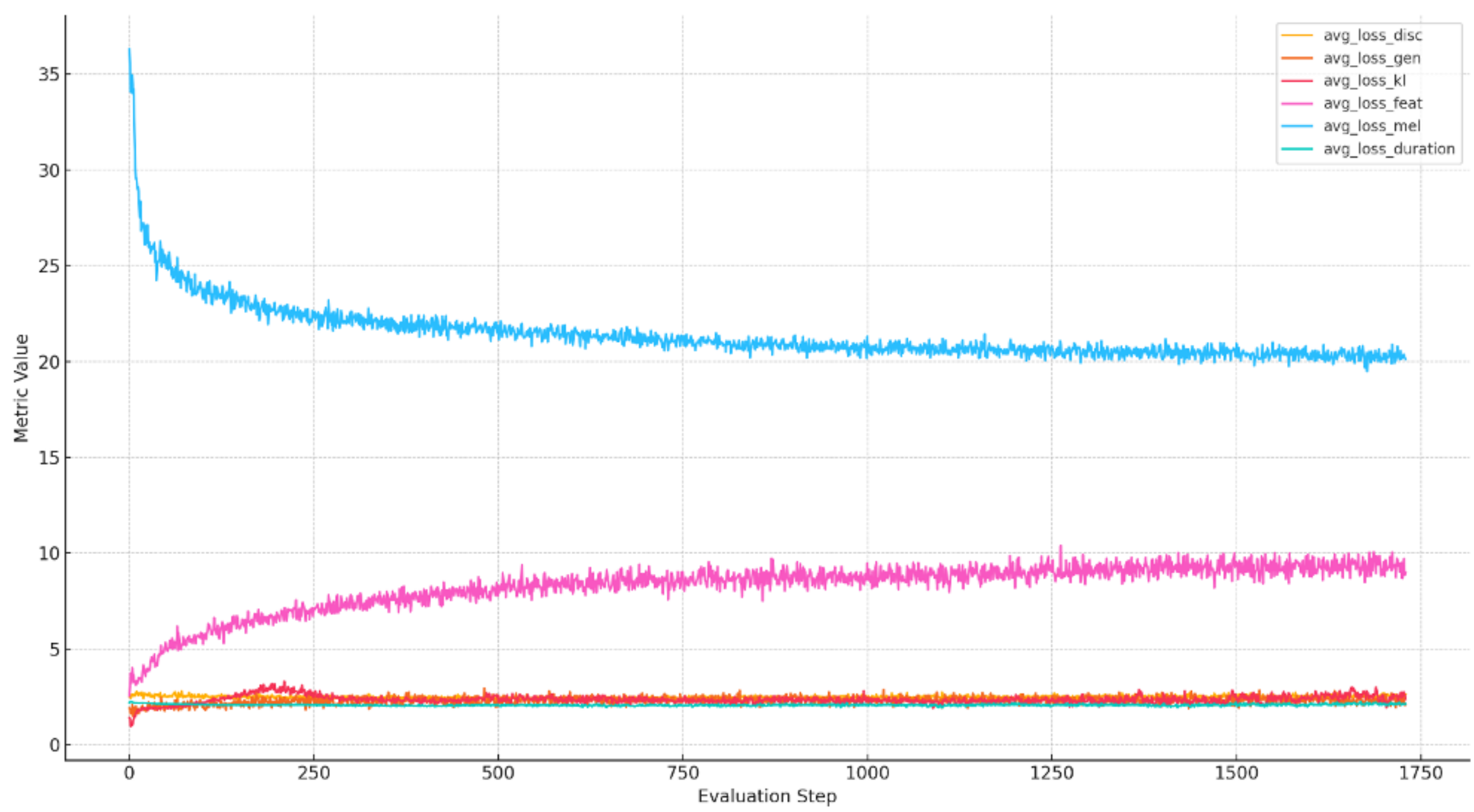

**Figure 10.0: YoruTTS-1.5 Training Performance Plots**

Furthermore, the **YoruTTS-1.5** model's checkpoint were saved to carry out inference with Yorùbá transcripts after training the model for 540 epochs and 1,000 epochs respectively. The Fundamental Frequency Root Mean Square Error (F0-RMSE) was adopted to evaluate the accuracy of the predicted pitch (F0) by **YoruTTS-1.5** compared to the reference real-time audio similar to Wu and King (2016). As shown in Figure 11.0, the F0-RMSE after 540 epochs is 72.85Hz while it is 63.54Hz after 1,000 epochs. However, there is no substantial improvement after the 1,000 epochs. Thus, we selected the checkpoints at 1,000 epochs as the acceptable **YoruTTS-1.5** model for inference. Although the F0-RMSE value of 63.542Hz appears a little bit high, the subjective assessment carried out using human evaluators indicates perceptual naturalness for all synthesized audios with **YoruTTS-1.5** model. This suggests that pitch deviations may not absolutely align with perceptual salience and other acoustic features like rhythm and timbre.

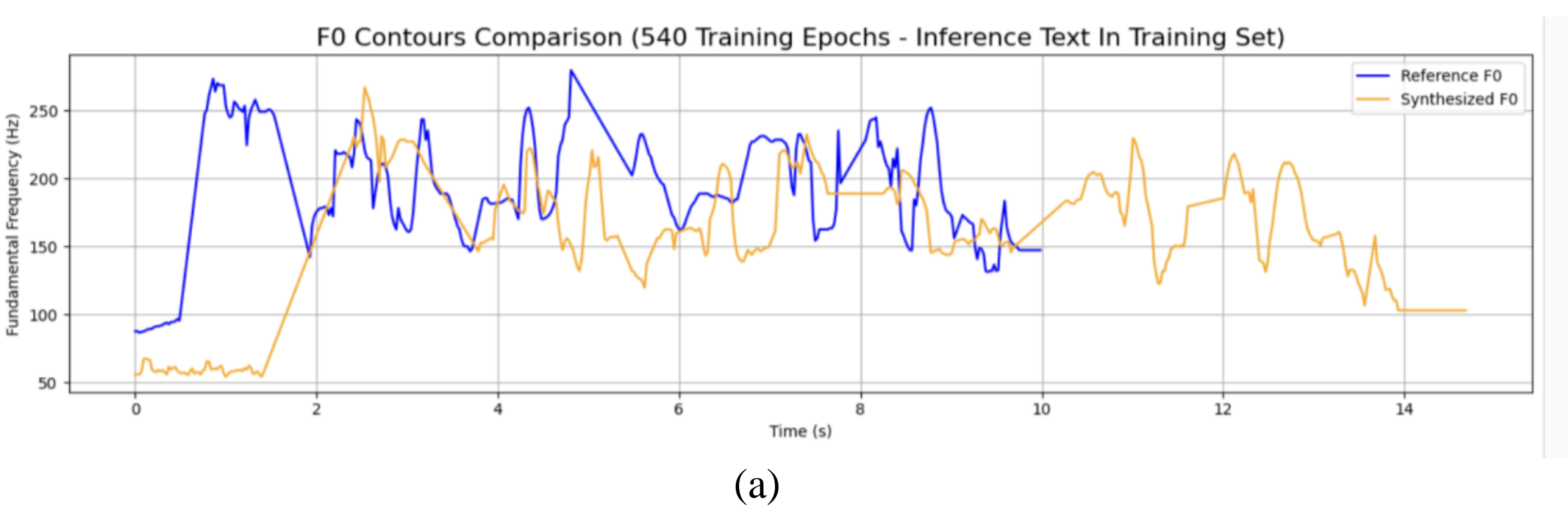

(a)

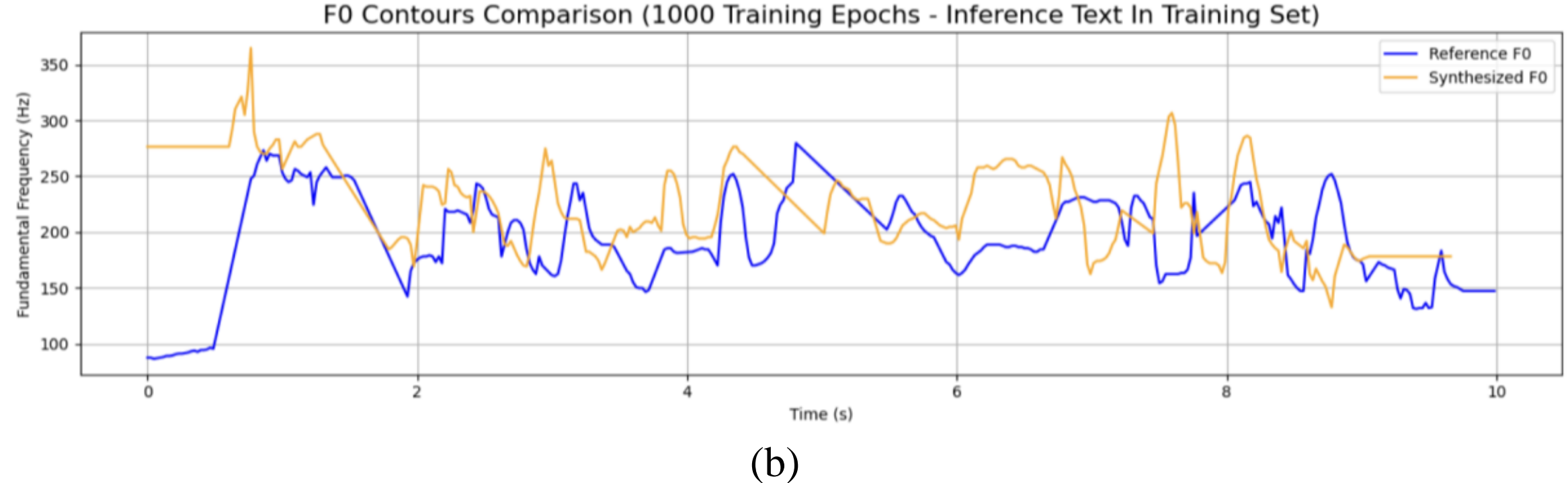

(b)

**Figure 11.0: F0-RMSE Comparison of YoruTTS-1.5 Synthesized Speech and Reference a) 540 Training Epochs, b) 1,000 Training Epochs**

## 5.0 Conclusion

This paper has introduced BENYO-S2ST-Corpus-1, a new bilingual English-to-Yorùbá direct speech-to-speech translation corpus designed to address the resource scarcity in this domain. The corpus advances existing efforts by including carefully curated parallel audio samples in the two languages, which preserves prosody and natural speech variation that are essential for tonal Yorùbá language. It enables end-to-end training and evaluation of direct S2ST systems as well as other speech processing tasks such as TTS, ASR and S2T. By providing a foundation for training and evaluating models in low-resource settings, this work contributes significantly to the broader goals of language inclusivity and equitable AI. Our future work will focus on leveraging agentic AI for the data curation architecture towards full automation. We will also expand the corpus to include more high-to-low resource language pairs with broader speaker variations to support multitask and multimodal speech research. Ultimately, BENYO-S2ST-Corpus is poised to become a critical resource for developing inclusive speech translation technologies for underrepresented languages in Africa and other regions.

## Acknowledgement

The first author, EA, is grateful to Google for financing this research through the Google Academic Research Award (GARA), which he received in 2024. The authors acknowledge Heather Cole-Lewis (Google) for her mentorship through the Google Academic Research Award (GARA). Google provided funding and mentoring but was not involved in data processing, model development, or the open-source code release. The Covenant University's Covenant Applied Informatics and Communication Africa Centre of Excellence (CApIC-ACE) funded through World Bank from 2019 to 2025 at Covenant University, Ota, Nigeria, is also acknowledged for the FEDGEN CodingHub (https://codinghub.fedgen.net) and HPC Cluster(https://fedgen.net/about/, https://bit.ly/3TB40s1) utilised for all the high performance computations done in this work.

## Appendix

**Table 1.0: Overview of Existing TTS Models Across Different Architecture Categories**

| S/N | Model | Architecture Category | Vocoder Used | Supported Languages | Reference |
|---|---|---|---|---|---|
| 1 | Tacotron | Autoregressive | Griffin-Lim | English | Wang et al., 2017 |
| 2 | Tacotron 2 | Autoregressive | WaveNet | English | Shen et al., 2018 |
| 3 | Char2Wav | Autoregressive | WaveNet | English | Sotelo et al., 2017 |
| 4 | Glow-TTS | Flow-Based (Non-Autoregressive) | HiFi-GAN | English | Kim et al., 2020 |
| 5 | VITS | Flow-Based (Non-Autoregressive) | Integrated (GAN + Flow) | Multilingual (via finetuning) | Kim et al., 2021 |
| 6 | YourTTS | Flow-Based (Zero-Shot, Non-Autoregressive) | Integrated (like VITS) | English, Portuguese, French | Casanova et al., 2022 |
| 7 | StyleTTS 2 | Diffusion-Based (Zero-Shot, Non-Autoregressive) | HiFi-GAN | English | Tian et al., 2023 |
| 8 | Diff-TTS | Diffusion-Based (Non-Autoregressive) | Parallel WaveGAN | English | Popov et al., 2021 |
| 9 | FastSpeech | Parallel Feedforward (Non-Autoregressive) | MelGAN, Parallel WaveGAN | English | Ren et al., 2019 |
| 10 | FastSpeech 2 | Parallel Feedforward (Non-Autoregressive) | HiFi-GAN | English | Ren et al., 2020 |
| 11 | FastSpeech 2s | Parallel Feedforward (Non-Autoregressive) | HiFi-GAN | English | Ren et al., 2022 |
| 12 | TalkNet | Parallel Feedforward (Non-Autoregressive) | HiFi-GAN | English | Beliaev et al., 2021 |
| 13 | ParaNet | Parallel Feedforward (Non-Autoregressive) | WaveGlow | English | Peng et al., 2019 |
| 14 | SpeedySpeech | Parallel Feedforward (Non-Autoregressive) | HiFi-GAN | English | Ren et al., 2021 |
| 15 | E2TTS | Parallel Feedforward (Non-Autoregressive) | HiFi-GAN | English | Eskimez et al., 2024 |
| 16 | XTTSv1 | Prompt-Based (Zero-Shot, Non-Autoregressive) | Integrated | Multilingual (16 languages) | Coqui.ai, 2023 |
| 17 | XTTS v2 | Prompt-Based (Zero-Shot, Non-Autoregressive) | Integrated | Multilingual (17 languages) | Coqui.ai, 2024 |
| 18 | META AI MMS | Prompt-Based (Zero-Shot, Non-Autoregressive) | Integrated | Multilingual (1000+ languages) | Babu et al., 2023 |